Anomalous Nernst effect in amorphous Tb-Fe-Co thin films

Hiroto Imaeda[1], Tsunehiro Takeuchi[1], Hiroyuki Awano[1], and Kenji Tanabe[1*]

[1]Toyota Technological Institute, Nagoya, 468-8511, Japan

*electronic mail: tanabe@toyota-ti.ac.jp

**Abstract**

We conducted a comprehensive study on the compositional dependence of the anomalous Nernst effect (ANE) in amorphous (amo.) Tb-Fe-Co thin films. The anomalous Nernst coefficient strongly depends not only on the Tb composition but also on the transition metal composition, reaching a maximum of 1.8 µV/K for amo. $Tb_{11.0}(Fe_{50.0}Co_{50.0})_{89.0}$. By evaluating the electrical and thermoelectric properties, it was clarified that this maximum is achieved by the superposition of two large contributions: $S_1$ arising from direct transverse electron conduction due to a temperature gradient, and $S_2$ resulting from the combined Seebeck and anomalous Hall effects. We discovered that the anomalous Nernst conductivity, which is attributed to Berry curvature, varied significantly with the transition metal, even in an amorphous material lacking long-range crystalline order. Our research indicates that it is possible to control the electronic states that influence thermoelectric properties, even in the amorphous state.

**1. Introduction**

Thermoelectric conversion is a phenomenon that directly converts heat into electricity, enabling the utilization of waste heat and precise thermal control. The most famous thermoelectric phenomenon is the Seebeck effect (SE). The SE is a phenomenon in which a thermoelectromotive force is generated when a temperature gradient is applied to a metal or semiconductor. Since the generated thermoelectromotive force is parallel to the temperature gradient, the SE is a longitudinal thermoelectric phenomenon. In contrast, the field of spin-caloritronics [1] has drawn attention to the anomalous Nernst effect (ANE), a transverse thermoelectric phenomenon [2-4]. The ANE is a phenomenon in which, when a temperature gradient is applied to a magnetic material, a thermoelectromotive force is generated in a direction perpendicular to both the temperature gradient and magnetization. It is expressed by the following equation:

$$\boldsymbol{E} = S_{\text{ANE}} \boldsymbol{\nabla} T \times \left(\frac{\boldsymbol{M}}{|\boldsymbol{M}|}\right) \quad (1)$$

where $\boldsymbol{E}$, $\boldsymbol{\nabla}T$, $S_{\text{ANE}}$, and $\boldsymbol{M}$ are electric field, temperature gradient, the anomalous Nernst coefficient, and magnetization, respectively. As $\boldsymbol{E}$ and $\boldsymbol{\nabla}T$ are orthogonal, the device must be elongated in the direction of the thermoelectromotive force (in-plane direction) to increase the voltage, as shown in Fig. 1. Conversely, the film thickness does not affect the voltage magnitude, eliminating the need for thick elements and facilitating thin film fabrication. This feature enables flexible device applications and the creation of devices with low thermal resistance.

Historically, the ANE was considered to be proportional to the saturation magnetization of ferromagnetic materials [5-7]. In 2017, Ikhlas *et al.* observed a relatively large $S_{\text{ANE}}$ of 0.6 µV/K at room temperature in the antiferromagnet $Mn_3Sn$, which has an extremely small saturation magnetization, comparable to ferromagnets like Fe and Co [8]. The origin of this ANE was attributed to a large effective magnetic field in k-space, i.e., Berry curvature, proposing a new guideline for materials development. This report triggered a rapid acceleration of research on the ANE originating from Berry curvature [9-16].

The anomalous Hall effect (AHE) is also closely related to Berry curvature. The AHE is a phenomenon where a voltage is generated perpendicular to both the current and magnetization directions when current is applied to a magnetic material. The AHE consists of an intrinsic contribution from Berry curvature and extrinsic contributions from spin-scattering effects (skew scattering [17] and side-jump scattering [18]). Skew scattering is the asymmetric scattering of spins due to spin-orbit interaction with impurities. Side-jump scattering is a phenomenon where the wave packet shifts sideways without a change in velocity due to spin-orbit interaction with impurities.

Recently, Berry curvature-driven AHE has been reported in amorphous structures such as Fe-Si [19], Fe-Ge [20], and $Co_2MnGa$ [21]. Zhao *et al.* found a giant AHE in amorphous $Co_2MnGa$ [21], and Fujiwara *et al.* reported a large ANE in amorphous Fe-Sn alloys [22]. These origins are interpreted as the existence of short-range crystalline fragments with large Berry curvature despite lacking long-range order. Amorphous materials have excellent characteristics for application, such as not requiring an annealing process and having high substrate selectivity.

Recently, heat flux sensors have garnered attention as an application for ANE [23]. For ANE-type heat flux sensors, a meander structure combining materials with different signs of $S_{\text{ANE}}$, as shown in Fig. 1, has been proposed. This structure is designed to generate a large voltage in response to a small heat flow. Research on ANE-type heat flux sensors is progressing not only toward high sensitivity but also toward achieving high coercive force for zero-magnetic-field operation [24]. In our previous work, we proposed that high sensitivity could be achieved with a 3D structure and demonstrated this using amorphous Tb-Co alloys thin films [25]. For this 3D structure, large perpendicular magnetic anisotropy (PMA) is crucial for high coercive force.

Rare-earth (RE) transition-metal (TM) alloy ferrimagnets such as Tb-Fe-Co alloys are known for exhibiting PMA and very large coercive force near the magnetic compensation point (MCP) [26]. The magnetic moments of the RE and TM elements couple antiferromagnetically. In the regions with less RE doping than the MCP, the TM element dominates the magnetic moment

(TM-rich), while in regions with more RE doping, the RE element dominates (RE-rich). Near the MCP, the net magnetization becomes very small, reducing the influence of the demagnetization field. This characteristic of having large PMA despite an amorphous structure is excellent for a 3D ANE-type heat flux sensor application.

Previous studies have reported on the ANE of various RE-TM alloys. Ando et al. [27] reported $S_{\text{ANE}}$ of 0.28 (0.08) μV/K for AlN (25 nm)/ amo. $Tb_{26(36)}(FeCo)_{74(64)}$ (20 nm)/ AlN (10 nm). It was found that the sign of $S_{\text{ANE}}$ reversed between the TM-rich amo. $Tb_{26}(FeCo)_{74}$ and the RE-rich amo. $Tb_{36}(FeCo)_{64}$. Seki et al. [28] investigated the anomalous Ettingshausen effect (AEE), the inverse effect of ANE, in Al (4 nm)/ amo. $Gd_{22}Co_{78}$ (30 nm)/ Al (4 nm). They observed AEE near the MCP and estimated $S_{\text{ANE}} = 0.18$ μV/K. Liu et al. [29] investigated the ANE of Ta (1 nm)/ Pt (3 nm)/ amo. Gd-Co (4.7 nm, 6.7 nm)/ Ta (3 nm) and reported $S_{\text{ANE}}$ of 0.15 and $-0.13$ μV/K for amo. $Gd_{16}Co_{84}$ (4.7 nm) and amo. $Gd_{26}Co_{74}$ (6.7 nm), respectively. They also suggested that the ANE of amo. Gd-Co might be related to the magnetic properties of the TM element rather than the net magnetization. Kobayashi et al. [30] systematically studied the Gd composition dependence of ANE in SiN (60 nm)/ amo. Gd-Fe (20 nm)/ SiN (5 nm), reporting a maximum $S_{\text{ANE}}$ of about 0.3 μV/K. More recently, Odagiri et al. [26, 31] investigated the systematic RE composition dependence of ANE in SiN (10 nm)/ amo. Gd(Tb)-Co (20 nm)/ SiN (3 nm). They clarified that the ANE properties hardly change with the RE element type, with $S_{\text{ANE}} > 1.0$ μV/K. They also reported that due to the low thermal conductivity $\kappa$ of amorphous materials, the material sensitivity for heat flux sensors ($S_{\text{ANE}}/\kappa$) becomes very large, making it a suitable material for heat flux sensors [31].

Although various studies have been conducted on RE-TM alloy samples, a problem is that $S_{\text{ANE}}$ cannot be directly compared when the buffer and cap layers are metallic, due to effects like shunting. A comparison of reports that use SiN for buffer and cap layers to eliminate this problem reveals that $S_{\text{ANE}}$ does not change between amo. Gd-Co [31] and amo. Tb-Co [26], but a large difference is reported between amo. Gd-Co [31] (max $S_{\text{ANE}} \approx 1.3$ μV/K) and amo. Gd-Fe [30] (max $S_{\text{ANE}} \approx 0.3$ μV/K). These results suggest that $S_{\text{ANE}}$ in RE-TM alloy ferrimagnets does not change significantly with RE element type but may vary greatly with the TM element type. However, no comprehensive study has been conducted to date that also includes the composition dependence of Fe-Co transition metal alloys.

In this study, we systematically investigated the compositional dependence of ANE in amorphous Tb-Fe-Co and clarified that the magnitude of $S_{\text{ANE}}$ differs significantly depending on the TM element. By examining the electrical resistivity, Hall resistivity, Seebeck coefficient, and anomalous Nernst coefficient, we derived the anomalous Nernst conductivity, which is attributed to Berry curvature, and found that it changes with the TM element. This result suggests that even in an amorphous structure lacking long-range order, the electronic states near the Fermi level of the TM element are strongly influenced by Berry curvature.

## 2. Experimental method

$SiN_z$ (10 nm)/ amo. $Tb_x(Fe_yCo_{100-y})_{100-x}$ (20 nm)/ $SiN_z$ (3 nm) thin films were deposited on $SiO_2$ glass substrates using RF and DC magnetron sputtering systems. Amorphous Tb-Fe-Co alloy thin films were co-sputtered using Tb (99.9 % purity), Co (99.9 % purity), $Fe_{25}Co_{75}$ (99.9 % purity), $Fe_{50}Co_{50}$ (99.9 % purity), $Fe_{75}Co_{25}$ (99.9 % purity), and Fe (99.9 % purity) targets. $SiN_z$ was used as an oxidation protection layer. The base pressure was below $3 \times 10^{-5}$ Pa, and the sputtering pressure was $\sim 2 \times 10^{-1}$ Pa. Samples with a Hall bar structure (Fig. 3 (a)) were fabricated using a metal mask. Au (~100 nm)/ Cr (~30 nm) electrodes were deposited using a resistive heat evaporation system. The composition of amo. Tb-Fe-Co was measured using an electron probe microanalyzer (EPMA). The amorphous state of the thin films was confirmed by X-ray diffraction (XRD). At room temperature, the electrical resistivity was measured by the 4-probe method, the Hall resistivity was measured with a magnetic field applied perpendicular to the sample plane, the Seebeck coefficient was measured with heat applied in the in-plane direction, and the anomalous Nernst coefficient was measured with a magnetic field applied perpendicularly and a temperature gradient applied in-plane. Magnetic properties were measured using an alternating gradient magnetometer (AGM).

## 3. Results and discussion

Figures 2(a) and 2(b) show the saturation magnetization ($M_s$) and coercive force ($H_c$), respectively, as a function of Tb composition, measured by applying a magnetic field perpendicular to the plane using AGM. $M_s$ decreased as the Tb composition increased relative to the Fe-Co elements. This is because the magnetic moments of Tb and Fe-Co couple antiferromagnetically. At around 25 % Tb, the magnetization compensation composition was reached, and the magnetization became almost zero. This point is the MCP. As the Tb composition was further increased, the magnetization recovered. This result was a typical feature observed in RE-TM alloy ferrimagnets; the Fe-Co (Tb) element is dominant on the lower (higher) Tb-doped side of the MCP, which is TM-rich (RE-rich).

Next, to investigate the dependence on the transition metal type, Co in Tb-Co was partially substituted with Fe. As the Fe composition increased, the MCP shifted to the RE-rich side, reaching the most RE-rich side for Tb-($Fe_{75}Co_{25}$) (Fig. 2(c)). When more Fe was doped, forming Tb-Fe, the MCP shifted to the TM side. In other words, the larger the saturation magnetization in the TM sublattice, the more the MCP shifts to the RE-rich side. The compositional dependence of the MCP is determined by the compensation between the magnitude of the TM magnetization and the RE magnetization; thus, it can be interpreted that as the TM magnetization increases (as the TM is close to $Fe_{75}Co_{25}$), the amount of RE required to bring the net magnetization to zero also increases.

As shown in Fig. 3(b), a temperature gradient ($\nabla T$) was applied in the in-plane direction (y-direction), a magnetic field ($H$) was applied in the perpendicular direction (z-direction), and the voltage ($V$) was measured in the direction perpendicular to both (x-direction). Figure 3(c) shows the relationship between the magnetic field and voltage for amo. $Tb_{18.3}(Fe_{50.0}Co_{50.0})_{81.7}$, where the color difference represents the temperature difference ($\Delta T$) across the sample. As $\Delta T$ increased, $V$ also increased. Figure 3(d) is a plot of the saturated $V$ values from Fig. 3(c) converted to electric field ($E$) on the x-axis, against $\nabla T$ (converted from $\Delta T$) on the y-axis. This figure shows that $E$ is proportional to $\nabla T$. The slope is $S_{ANE}$, which was calculated by a linear fit. Using this derivation method, $S_{ANE}$ for amo. $Tb_x(Fe_yCo_{100-y})_{100-x}$ ($y$ = 0, 25, 50, 75, 100) were determined (Fig. 3(e)). For $y$ = 0, 25, 50, 75, the TM-rich samples had $S_{ANE} > 0$ and the RE-rich samples had $S_{ANE} < 0$. However, for amo. $Tb_xFe_{100-x}$ ($y$ = 100), $S_{ANE} < 0$ when $x = 0$ and $S_{ANE} > 0$ when $x > 0$, showing a different result from $y$ = 0, 25, 50, 75 cases. Except for amo. $Tb_xFe_{100-x}$ ($y$ = 100), as Tb was doped into TM, the absolute value of $S_{ANE}$ increased, reached a maximum within the TM-rich region, and then decreased. Also, the Tb composition at which $S_{ANE}$ peaked decreased as the Fe composition increased. Amorphous $Tb_xFe_{100-x}$ ($y$ = 100) had a very small $S_{ANE}$ regardless of the Tb composition. The maximum $S_{ANE}$ was 1.8 μV/K for amo. $Tb_{11.0}(Fe_{50.0}Co_{50.0})_{89.0}$. This value is comparable to the 2.0 μV/K of amo. $Fe_{87}Sn_{13}$, which is known to have a very large $S_{ANE}$ for an amorphous material [22].

The results obtained for amo. $Tb_xCo_{100-x}$ ($y$ = 0) and amo. $Tb_x(Fe_{50.0}Co_{50.0})_{100-x}$ ($y$ = 50) show roughly the same trend as previous studies [26, 32], but the MCP differs. Compared to our results, the MCP in Odagiri *et al*. [26] is shifted to the TM-rich side, while the MCP in Ando *et al*. [32] is shifted to the RE-rich side. Comparison of the base pressures of the magnetron sputtering systems reveals that they were less than $1.0 \times 10^{-5}$ Pa for Odagiri *et al*., less than $3 \times 10^{-5}$ Pa for our experiment, and $10^{-4}$ Pa for Ando *et al*. At lower vacuum levels, Tb oxidizes during deposition, losing its magnetic moment, thus increasing the amount of Tb required to bring the net magnetization to zero. Although the MCP changed depending on the fabrication conditions, there is almost no difference in the magnitude of $S_{ANE}$. This finding indicates that it is possible to produce different signs of $S_{ANE}$ for the same composition by adjusting the base pressure.

Figure 3(f) and (g) show the TM composition dependence of $S_{ANE}$ for amo. $Tb_{\sim 20}(Fe_yCo_{100-y})_{\sim 80}$ and amo. $Tb_{\sim 10}(Fe_yCo_{100-y})_{\sim 90}$, respectively. For amo. $Tb_{\sim 20}(Fe_yCo_{100-y})_{\sim 80}$, $S_{ANE}$ increased as Fe was doped from $y$ = 0, but decreased when $y > 25$, and became very small at $y$ = 100. Interestingly, this trend is very similar to that of amo. Sm-Fe-Co, which is the ferromagnetic RE-TM alloy, reported by Modak *et al*. [33]. This experimental fact suggests that the magnitude of $S_{ANE}$ is almost independent of the coupling direction of TM-RE magnetic moments. When the RE composition changed, as shown in Fig. 3(g), the trend for amo. Tb$_{\sim 10}$(Fe$_y$Co$_{100-y}$)$_{\sim 90}$ differed from that of amo. Tb$_{\sim 20}$(Fe$_y$Co$_{100-y}$)$_{\sim 80}$. In this case, $S_{ANE}$ shows a maximum value at $y = 50$.

Figure 4 shows the electrical properties of each sample. As shown in Fig. 4(a), the electrical resistivity ($\rho_{xx}$) was derived by measuring the voltage in the x-direction while applying current in the x-direction (Fig. 4(b)). Regardless of the TM, $\rho_{xx}$ increased monotonically with Tb-doping. The Hall resistivity ($\rho_{xy}$) was measured in the configuration shown in Fig. 4(c), by applying a current in the x-direction and a magnetic field in the z-direction, and measuring the voltage in the y-direction. Figure 4(d) and (e) show the relationship between the magnetic field and Hall resistivity for TM-rich and RE-rich amo. Tb$_x$(Fe$_{50.0}$Co$_{50.0}$)$_{100-x}$ ($y = 50$), respectively. The color difference represents the difference in Tb composition. It can be confirmed that the coercive force increased as the composition approached the MCP. Figure 4(f) shows the Hall resistivity versus Tb composition for each sample. Regardless of the TM, the absolute value of $\rho_{xy}$ increased with Tb doping, peaked around $y \approx 25$, and then decreased. Figure 4(g) shows the relationship between the Hall resistivity and TM composition. Both TM-rich (Tb $\approx$ 10) and RE-rich (Tb $\approx$ 30) show a maximum at Tb$_x$(Fe$_{75.0}$Co$_{25.0}$)$_{100-x}$ ($y = 75$), and the value decreased as the composition moved away from $y = 75$. Interestingly, the trend in the magnitude of the Hall resistivity is the same as the trend in the magnitude of the TM saturation magnetization; that is, in Fe-Co alloys, saturation magnetization is maximum at Fe$_{75}$Co$_{25}$.

We investigated the details of the electron transport properties using the scaling relation proposed by Onoda *et al*. [34], which divides the properties into the dirty, intrinsic, and clean regions using electrical conductivity ($\sigma_{xx}$) and anomalous Hall conductivity ($\sigma_{xy}$) (Fig. 4(h)). These values were derived using the equations $\sigma_{xx} = \frac{\rho_{xx}}{\rho_{xx}^2+\rho_{xy}^2}$ and $\sigma_{xy} = -\frac{\rho_{xy}}{\rho_{xx}^2+\rho_{xy}^2}$. From Fig. 4(h), amo. Tb$_x$(Fe$_y$Co$_{100-y}$)$_{100-x}$ is distributed in the dirty and intrinsic regions depending on the composition, entering the dirty region as the Tb composition increases. Although a power law has been proposed [34], this relationship was not observed in the plot of Tb composition dependence. It is thought that the respective contributions to the origin of the AHE change with Tb composition. A detailed analysis would require temperature dependent results.

Next, the thermoelectric properties were investigated (Fig. 5). $S_{SE}$ was derived by applying a temperature gradient in the direction shown in Fig. 5(a) and measuring the voltage (Fig. 5(b)). Except for amo. Tb$_x$Fe$_{100-x}$ ($y = 100$), $S_{SE}$ showed negative values, and its absolute value decreased with the amount of Tb doping. On the other hand, for amo. Tb$_x$Fe$_{100-x}$ ($y = 100$), the sign of $S_{SE}$ was reversed with Tb doping in Fe.

$S_{\text{ANE}}$ can be expressed using $\rho_{xx}$, $\rho_{xy}$, $S_{\text{SE}}$, and the anomalous Nernst conductivity $\alpha_{xy}$ by following equation (2):

$$S_{\text{ANE}} = \rho_{xx}\alpha_{xy} + S_{\text{SE}}\frac{\rho_{xy}}{\rho_{xx}} \qquad (2)$$

where the first term on the right side, $\rho_{xx}\alpha_{xy}$, and the second term, $S_{\text{SE}}\frac{\rho_{xy}}{\rho_{xx}}$, are respectively $S_1$, which directly converts the temperature gradient to a transverse current, and $S_2$, which is a combination of the Seebeck effect and anomalous Hall effect. From Eq. (2), $|\alpha_{xy}|$ was calculated using the measured $S_{\text{ANE}}$, $\rho_{xx}$, $\rho_{xy}$, and $S_{\text{SE}}$ (Fig. 5(c)). Regarding the RE composition dependence, $|\alpha_{xy}|$ tended to decrease with increasing Tb composition. On the other hand, regarding the TM composition dependence, $|\alpha_{xy}|$ was maximum at Co ($y = 0$), decreased as Fe was doped, and was minimum at Fe$_{75.0}$Co$_{25.0}$ ($y = 75$). $\alpha_{xy}$ is considered to be derived from Berry curvature, and it varies systematically with TM, even in an amorphous material without long-range order. This suggests that although the material is amorphous, the Fermi level could be modulated by the number of electrons in the TM.

Figure 6(a) and (b) show triangular contour plots of $|\alpha_{xy}|$ and the Hall angle $|\sigma_{xy}/\sigma_{xx}|$ for amo. Tb-Fe-Co composition, respectively. $|\alpha_{xy}|$ was maximum when TM was Co and minimum when TM was Fe$_{75.0}$Co$_{25.0}$. In contrast, $|\sigma_{xy}/\sigma_{xx}|$ was maximum when TM was Fe$_{75.0}$Co$_{25.0}$ and minimum when TM was Co, showing completely different behavior from $|\alpha_{xy}|$. We considered this using the Mott relation [35, 36]:

$$\alpha_{xy} \approx -\frac{\pi^2}{3}\frac{k_B^2 T}{|e|}\left[\frac{\partial \sigma_{xy}(\varepsilon, T)}{\partial \varepsilon}\right]_{\varepsilon = E_F} \qquad (3)$$

where $k_B$ is the Boltzmann constant, $e$ is the elementary charge, $\varepsilon$ is energy, $E_F$ is the Fermi energy, and $\sigma_{xy}(\varepsilon, T)$ is the spectral conductivity. The Mott relation is generally applied to materials with long-range order and discussed based on band structure, and thus cannot be applied to amorphous materials. However, previous studies have reported the influence of short-range crystalline fragments [21, 22], and assuming that amorphous Tb-Fe-Co alloys also contain short-range crystalline fragments, we can discuss this using the Mott relation. In this case, the difference between $|\alpha_{xy}|$ and $|\sigma_{xy}/\sigma_{xx}|$ can be interpreted as a difference in the energy at which the absolute value of $\sigma_{xy}$ is large and the energy at which the energy derivative of $\sigma_{xy}(\varepsilon, T)$ at $\varepsilon = E_F$ is large, with the Fermi energy being modulated by the number of TM electrons.

Figure 7(a) shows the Tb composition dependence of $S_{\text{ANE}}$, $S_1$, $S_2$ for amo. Tb$_x$(Fe$_y$Co$_{100-y}$)$_{100-x}$. Regardless of the TM, the signs of $S_1$ and $S_2$ were reversed at the MCP. For $y = 0, 25, 50, 75$, $S_1$ and $S_2$ had the same sign, while for $y = 100$, $S_1$ and $S_2$ had opposite signs and canceled each other out. This is the reason why $S_{\text{ANE}}$ for amo. Tb$_x$Fe$_{100-x}$ ($y = 100$) was very small. Figures 7(b)-(d) show the triangular contour plots of the absolute value of $S_{\text{ANE}}$, $S_1$, and $S_2$ for amo. Tb$_x$(Fe$_y$Co$_{100-y}$)$_{100-x}$, respectively. As shown in Fig. 7(b), $|S_{ANE}|$ was maximum at Tb$_{11.0}$(Fe$_{50.0}$Co$_{50.0}$)$_{89.0}$ and decreased as the composition moved away from this point. $|S_1|$ was large for amo. Tb$_x$Co$_{100-x}$ ($y = 0$) and small for amo. Tb$_x$(Fe$_{75}$Co$_{25}$)$_{100-x}$ ($y = 75$) (Fig. 7(c)). This trend matched that of $|\alpha_{xy}|$. $|S_2|$ was large for amo. Tb$_x$(Fe$_{75}$Co$_{25}$)$_{100-x}$ ($y = 75$) and small for amo. Tb$_x$Co$_{100-x}$ ($y = 0$) (Fig. 7(d)), which matched the trend of $|\rho_{xy}|$. It was clarified that the maximum $|S_{ANE}|$ is not due to $|S_1|$ or $|S_2|$ being maximum, but rather because both have the same sign and large values, and the maximum is achieved by their summation.

## 4. Conclusion

We systematically investigated the anomalous Nernst effect in amo. Tb-Fe-Co thin films. Regarding the TM composition dependence, $S_1$, which is the product of the anomalous Nernst conductivity and longitudinal resistivity, reached its maximum for amo. Tb-Co, while $S_2$, which is determined by the Seebeck and Hall effects, reached its maximum for amo. Tb$_x$(Fe$_{75}$Co$_{25}$)$_{100-x}$. $S_{ANE}$ was maximum at 1.8 μV/K for amo. Tb$_{11.0}$(Fe$_{50.0}$Co$_{50.0}$)$_{89.0}$, a large value. It was clarified that this maximum was achieved because both $S_1$ and $S_2$ had large values. The anomalous Nernst conductivity originating from Berry curvature varies significantly depending on the transition metal despite the absence of long-range order in amo. Tb-Fe-Co, suggesting it may be influenced by electronic states near the Fermi level. Furthermore, a large coercive force is necessary for practical applications requiring stability against external magnetic field. Amorphous Tb-Fe-Co has a very large coercive force near the MCP, and it was clarified that Tb composition at the MCP changes with TM composition. Comparison with previous research also indicated the possibility of modulating the MCP for the same composition by adjusting the deposition vacuum level. These results provide a valuable guideline for research aiming to achieve both a large anomalous Nernst coefficient and a high coercive force.

**ACKNOWLEDGMENTS**

This work was partially supported by JST A-STEP Grant Number JPMJTR25TA, the Fuji Science and Technology Foundation, and the Tanikawa Foundation.

**AUTHOR DECLARATIONS**

Conflict of Interest

The authors have no conflicts to disclose.

**Author Contributions**

**Hiroto Imaeda**: Conceptualization (equal); Data curation (lead); Formal analysis (lead); Investigation (lead); Writing – original draft (lead); Writing – review & editing (lead). **Tsunehiro Takeuchi**: Investigation (supporting); Writing – review & editing (supporting). **Hiroyuki Awano**: Investigation (supporting); Resources (equal); Writing – review & editing (supporting). **Kenji Tanabe**: Conceptualization (equal); Data curation (supporting); Formal analysis (supporting); Funding acquisition (lead); Investigation (supporting); Project administration (lead); Resources (equal); Supervision (lead); Writing – original draft (supporting); Writing – review & editing (supporting).

**DATA AVAILABILITY**

The data that support the findings of this study are available from the corresponding author upon reasonable request.

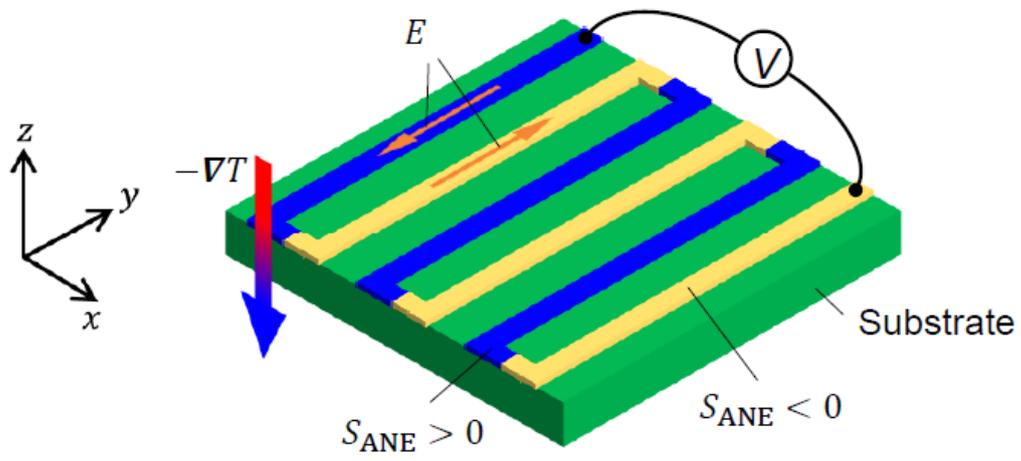

Fig. 1 Schematic diagram of a heat flux sensor based on the ANE with a meander structure. Blue and yellow wires indicate thermoelectric magnetic materials with $S_{\text{ANE}} > 0$ and $S_{\text{ANE}} < 0$, respectively.

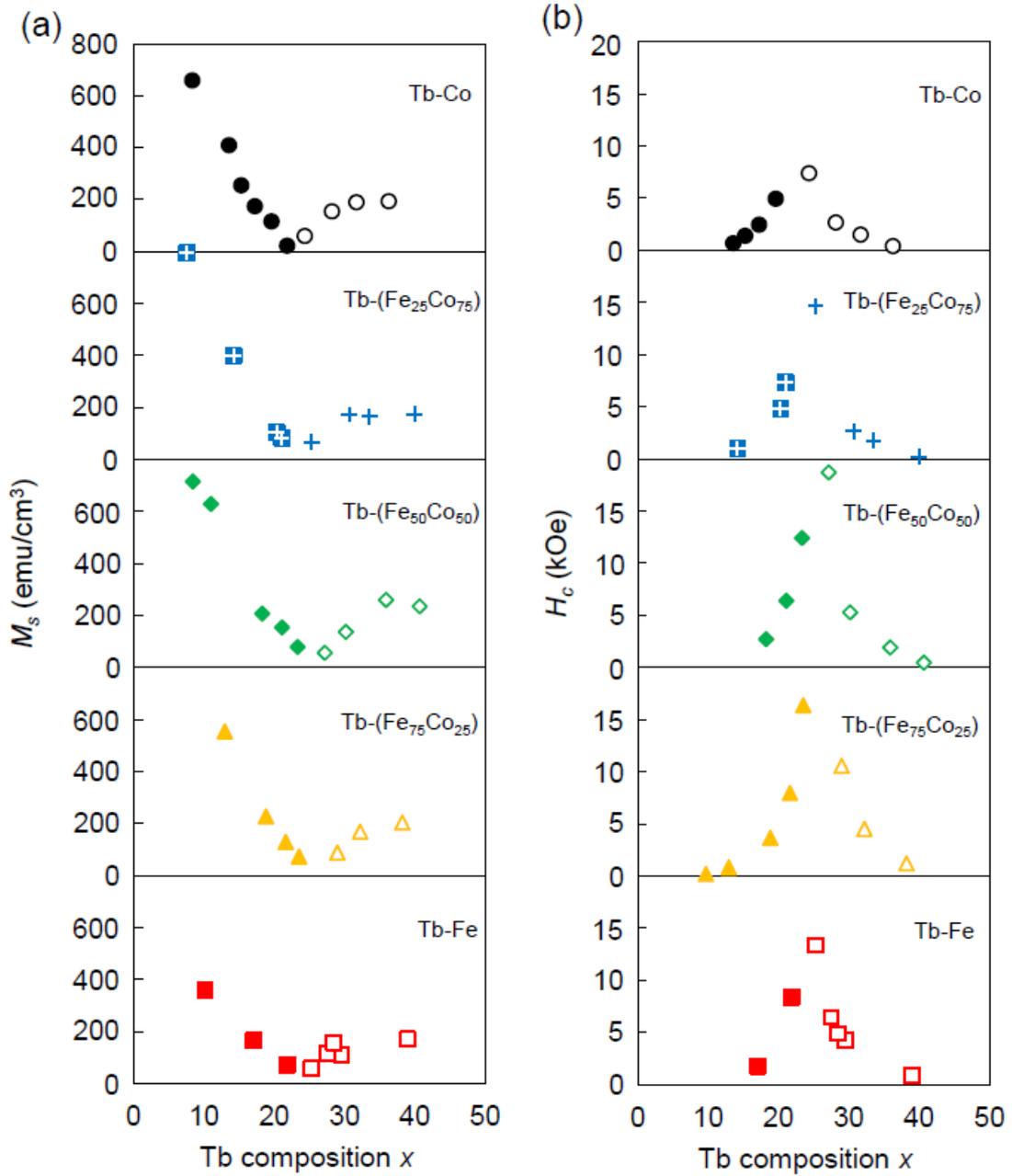
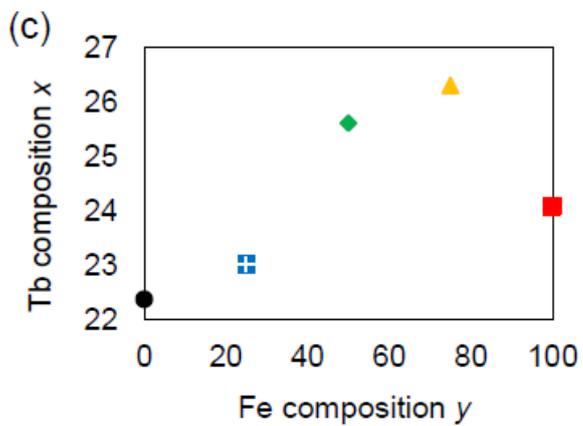

Fig. 2 (a-b) Tb composition dependences of saturation magnetization (a) and coercive force (b) of amo. $Tb_x(Fe_yCo_{100-y})_{100-x}$. Each color represents $y = 0$ for black circles, $y = 25$ for blue crosses, $y = 50$ for green rhombuses, $y = 75$ for yellow triangles, and $y = 100$ for red squares. The closed and open points correspond to TM-rich and RE-rich samples, respectively. (c) Tb compositions at magnetic compensation points for each Fe composition.

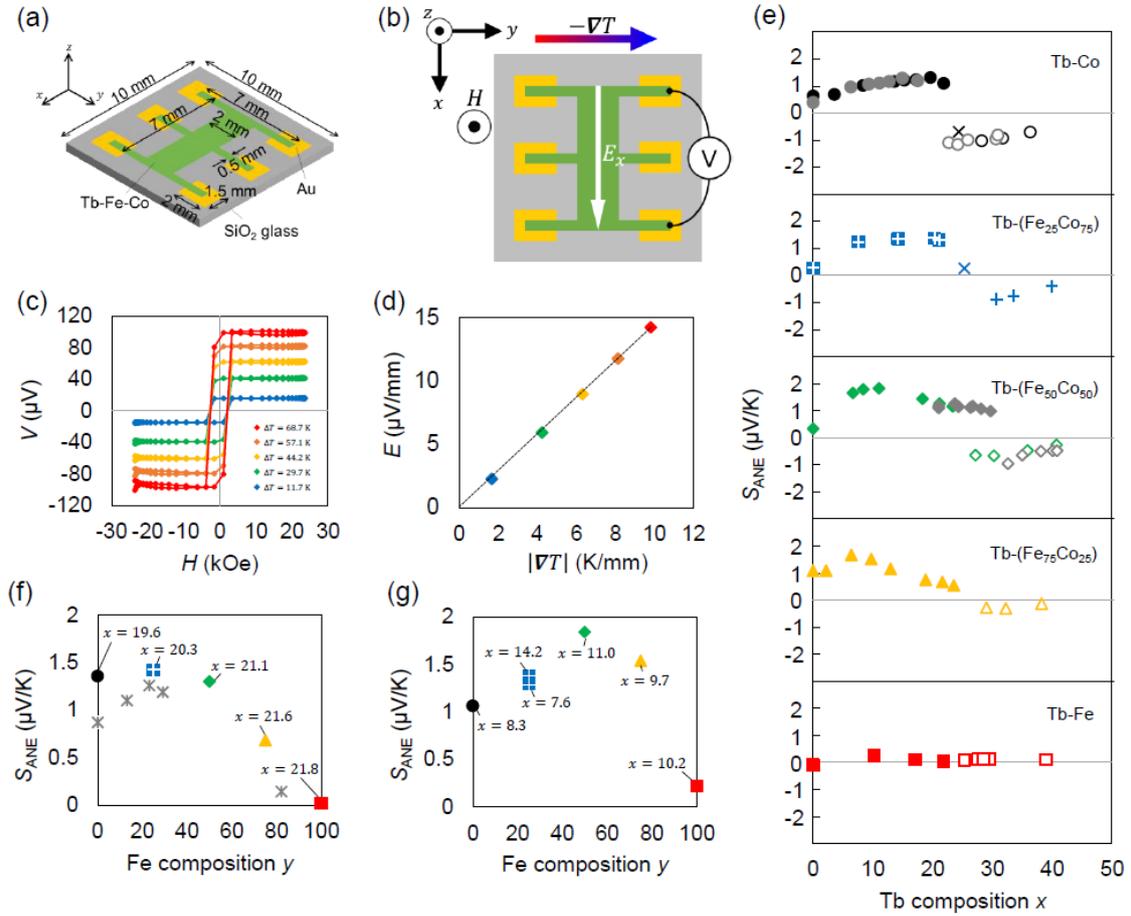

Fig. 3 (a) Schematic diagram of the sample structure. (b) Schematic illustration of setup for measuring the ANE. (c) Relationship between magnetic field and anomalous Nernst voltage of amo. $Tb_{18.3}(Fe_{50}Co_{50})_{81.7}$. The differences in color represent the differences in the temperature of the samples. (d) Relationship between temperature gradient and electric field derived from Fig. 2(c). The dotted line is the line fit and the slope corresponds to $S_{ANE}$. (e) Tb composition dependence of $S_{ANE}$. The color definitions are the same as in Fig. 2(a-b). Gray circles and rhombuses are the results of previous studies [31, 32]. (e-f) Transition metal composition dependence of $S_{ANE}$ in amo. $Tb_x(Fe_yCo_{100-y})_{100-x}$ at $x \approx 20$ (e) and $x \approx 10$ (f). The gray dots indicate $S_{ANE}$ in amo. Sm-Fe-Co from previous study [33].

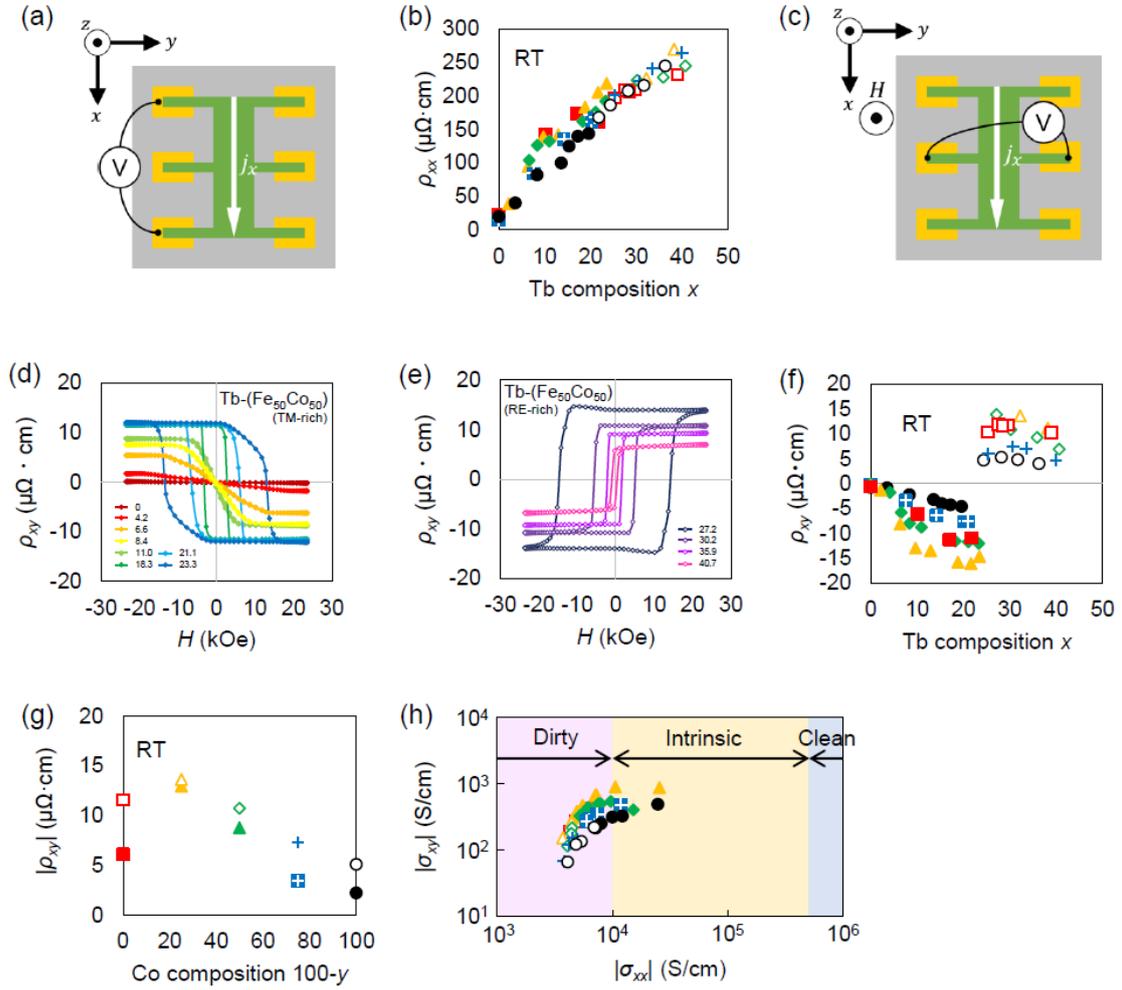

Fig. 4 (a) Schematic illustration of setup for measuring the electrical resistivity. (b) Electrical resistivity as a function of Tb composition. The color definitions are the same as in Fig. 2(a-b). (c) Schematic illustration of setup for measuring the Hall resistivity. (d-e) The magnetic field dependence of the Hall resistivity of amo. $Tb_x(Fe_{50.0}Co_{50.0})_{100-x}$, TM-rich (d) and RE-rich (e). Different colors represent different compositions of Tb. (f) Tb composition dependence of the Hall resistivity for different transition metal compositions. The color definitions are the same as in Fig. 2(a-b). (g) Transition metal dependence of the Hall resistivity at $x \approx 10$ (closed circles) and $x \approx 30$ (open circles). (h) The relationship between $|\sigma_{xy}|$ and $|\sigma_{xx}|$. The color definitions are the same as in Fig. 2(a-b). The purple, yellow, and blue areas represent the dirty, intrinsic, and clean regions, respectively.

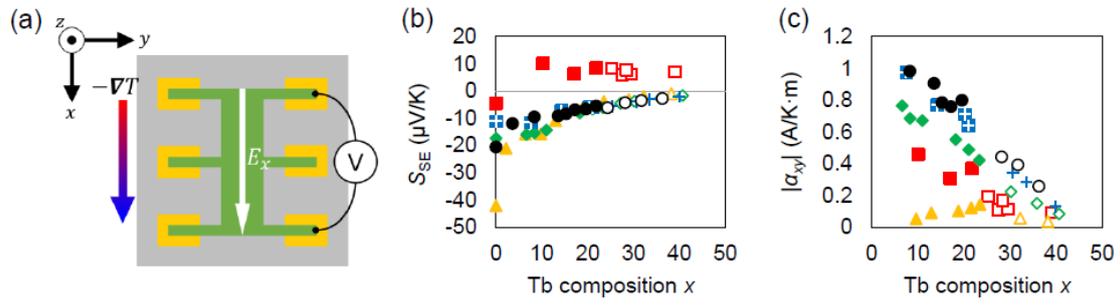

Fig. 5 (a) Schematic illustration of setup for measuring the SE. (b-c) Tb composition dependence of $S_{\text{SE}}$ (b) and $|\alpha_{xy}|$ (c) for each transition metal. The color definitions are the same as in Fig. 2(a-b).

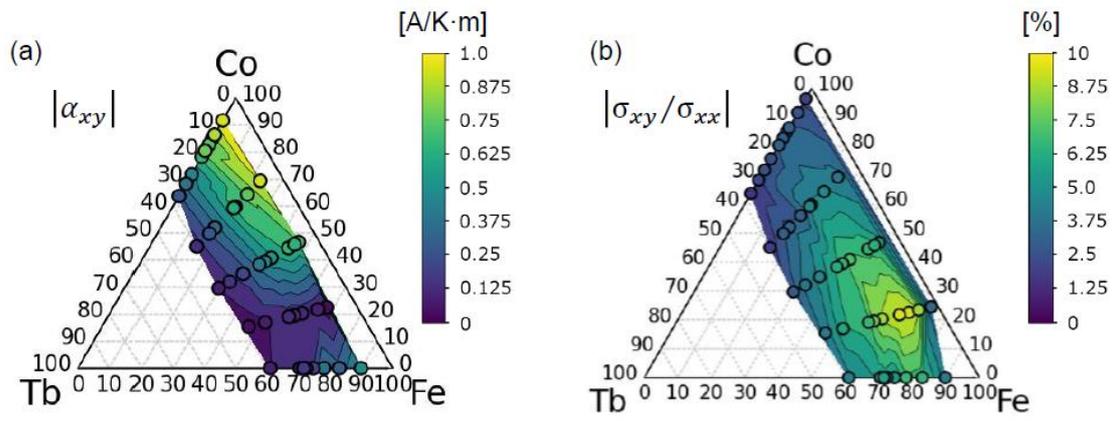

Fig. 6 (a-b) Triangular contour map of $|\alpha_{xy}|$ (a) and $|\sigma_{xy}/\sigma_{xx}|$ (b) for amo. Tb-Fe-Co composition.

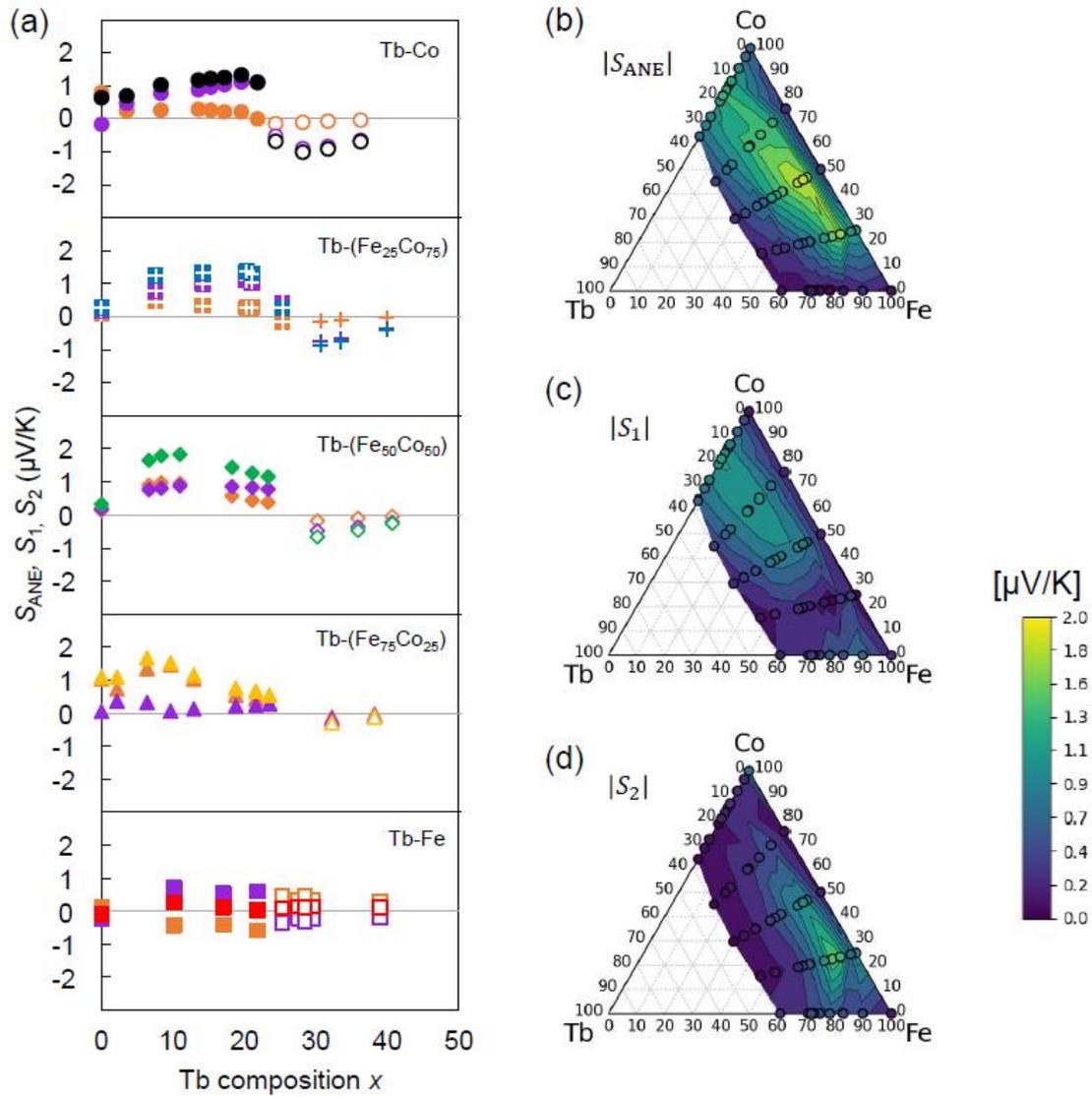

Fig. 7 (a) Tb composition dependence of $S_{ANE}$, $S_1$ and $S_2$. The color definitions are the same as in Fig. 2(a-b). The orange dots are $S_1$, and the purple dots are $S_2$. (b-d) Triangular contour map of $S_{ANE}$ (b), $S_1$ (c) and $S_2$ (d) for amo. Tb-Fe-Co composition.